\documentclass[sigconf]{acmart}

\usepackage{booktabs} % For formal tables
\usepackage{subcaption}
\usepackage{multirow}
\usepackage{enumitem}
\usepackage{array}
\usepackage{bm}
\usepackage{amsthm}
\usepackage{amsmath}
\newcolumntype{L}[1]{>{\raggedright\let\newline\\\arraybackslash\hspace{0pt}}m{#1}}
\newcolumntype{C}[1]{>{\centering\let\newline\\\arraybackslash\hspace{0pt}}m{#1}}
\newcolumntype{R}[1]{>{\raggedleft\let\newline\\\arraybackslash\hspace{0pt}}m{#1}}

\begin{document}
%\title{Temporal Network Embedding: A Neighborhood Formation Sequence Perspective}
%\title{Do Co-purchases Reveal Preferences? Explainable Recommendation with Attribute Networks}
\title{Recommendation with Attribute-aware Product Networks: A Representation Learning Model}

%\author{Yuan Zuo, Guannan Liu$^{1*}$, Hao Lin$^{1}$, Jia Guo$^{1}$, Xiaoqian Hu$^{1}$, Junjie Wu$^{1,2,3}$}
%\affiliation{%
%  \institution{$^1$School of Economics and Management, Beihang University, Beijing 100191, China}
%  \institution{$^2$Beijing Advanced Innovation Center for Big Data and Brain Computing, Beihang University, Beijing 100191, China}
%  \institution{$^3$Beijing Key Laboratory of Emergency Support Simulation Technologies for City Operations, Beihang University, Beijing 100191, China}
%  {\it $^*$corresponding author: liugn@buaa.edu.cn}
%}

\author{Guannan Liu}\authornote{Corresponding author}
\affiliation{
	\institution{Beihang University}
	%\city{Beijing 100191, China}
}
\email{liugn@buaa.edu.cn}

\author{Liang Zhang}
\affiliation{
	%\department{School of Economics and Management}
	\institution{Beihang University}
	%\city{Beijing 100191, China}
}
\email{zhangliang2017@buaa.edu.cn}

\author{Junjie Wu}
\affiliation{
	%\department{School of Economics and Management}
	\institution{Beihang University}
	%\city{Beijing 100191, China}
}
\email{wujj@buaa.edu.cn}

\author{Xiao Fang}
\affiliation{
	%\department{School of Economics and Management}
	\institution{University of Delaware}
	%\city{Beijing 100191, China}
}

%
%\author{Hao Lin}
%\affiliation{
%	\department{School of Economics and Management}
%	\institution{Beihang University}
%	\city{Beijing 100191, China}
%}
%\email{linhao2014@buaa.edu.cn}
%
%\author{Jia Guo}
%\affiliation{
%	\department{School of Economics and Management}
%	\institution{Beihang University}
%	\city{Beijing 100191, China}
%}
%\email{guojia1608@buaa.edu.cn}
%
%\author{Xiaoqian Hu}
%\affiliation{
%	\department{School of Economics and Management}
%	\institution{Beihang University}
%	\city{Beijing 100191, China}
%}
%\email{huxiaoqian@buaa.edu.cn}
%
%\author{Junjie Wu}
%\affiliation{
%	\department[0]{School of Economics and Management}
%	\department[1]{Beijing Advanced Innovation Center for Big Data and Brain Computing}
%%	\department[2]{Beijing Key Laboratory of Emergency Support Simulation Technologies for City Operations}
%	\institution{Beihang University}
%	\city{Beijing 100191, China}
%}
%\additionalaffiliation{
%	\department{Beijing Key Laboratory of Emergency Support Simulation Technologies for City Operations, }
%	\institution{Beihang University}
%	\city{Beijing 100191, China}
%}
%\email{wujj@buaa.edu.cn}

%\renewcommand{\shortauthors}{Y. Zuo et al.}

\begin{abstract}
With the prosperity of business intelligence, recommender systems have evolved into a new stage that we not only care about what to recommend, but why it is recommended. Explainability of recommendations thus emerges as a focal point of research and becomes extremely desired in e-commerce. Existent studies along this line often exploit item attributes and correlations from different perspectives, but they yet lack an effective way to combine both types of information for deep learning of personalized interests. In light of this, we propose a novel graph structure, \emph{attribute network}, based on both items' co-purchase network and important attributes. A novel neural model called \emph{eRAN} is then proposed to generate recommendations from attribute networks with explainability and cold-start capability. Specifically, eRAN first maps items connected in attribute networks to low-dimensional embedding vectors through a deep autoencoder, and then an attention mechanism is applied to model the attractions of attributes to users, from which personalized item representation can be derived. Moreover, a pairwise ranking loss is constructed into eRAN to improve recommendations, with the assumption that item pairs co-purchased by a user should be more similar than those non-paired with negative sampling in personalized view. Experiments on real-world datasets demonstrate the effectiveness of our method compared with some state-of-the-art competitors. In particular, eRAN shows its unique abilities in recommending cold-start items with higher accuracy, as well as in understanding user preferences underlying complicated co-purchasing behaviors.
\end{abstract}

%\begin{CCSXML}
%<ccs2012>
%<concept>
%<concept_id>10002951.10003227.10003351</concept_id>
%<concept_desc>Information systems~Data mining</concept_desc>
%<concept_significance>500</concept_significance>
%</concept>
%<concept>
%<concept_id>10002951.10002952.10002953.10010146.10010818</concept_id>
%<concept_desc>Information systems~Network data models</concept_desc>
%<concept_significance>300</concept_significance>
%</concept>
%<concept>
%<concept_id>10010147.10010257.10010258.10010260.10010271</concept_id>
%<concept_desc>Computing methodologies~Dimensionality reduction and manifold learning</concept_desc>
%<concept_significance>500</concept_significance>
%</concept>
%</ccs2012>
%\end{CCSXML}



\maketitle

\section{Introduction}
\label{sect:intro}

Recommender systems play indispensable roles in contemporary e-commerce by empowering consumers to reach their preferred products more efficiently~\cite{adomavicius2005toward}. Traditionally, recommendation methods rely on the similarity between items and recommend the most ``similar'' items in terms of users' historical purchasing records. Particularly, in item-based \emph{collaborative filtering} (CF)~\cite{sarwar2001item,kabbur2013fism}, the similarity between items can arise from the co-purchase relationships between items, \emph{i.e.}, two items would be regarded similar if they have been co-purchased by many other users in history.

The underlying factors that drive the co-purchases of items, however, may not be uniform for different users, and thus such co-purchases would not necessarily reveal users' genuine preferences. In reality, users may have their own desired feature aspects when considering to buy an item. For example, two movies may be co-watched by users, but some users may only be driven by the same leading actor, while others may be in favor of the same director. To infer users' underlying individual preferences from co-purchased items, however, have not been well addressed in prior studies. This is also closely related to the concept of\emph{explainable recommendations}, which has absorbed great research interests and become extremely desired in e-commerce~\cite{zhao2014we,zhang2018explainable}.  

In this paper, we aim at integrating items' co-purchase information with items' attribute information to deep learn users' personalized interests. Based on the item correlations formed in co-purchases, we propose a novel item network structure called \emph{attribute network}. An attribute network is essentially a customized sub-network of the item co-purchase network, formed by keeping only the connections between two items that have the given item attribute in both, {\it e.g.}, two movies with a same director. By decomposing a co-purchase network into various attribute networks, we can specify the diverse driving forces underlying the co-purchase network, which can then be used to characterize a user's individual preference and explain which aspects of a recommended item that he/she would desire most, {\it i.e.}, generate explainable recommendations. For those items that have never been purchased and thus cannot enter the co-purchase network, they can still enter attribute networks as long as they have attribute information, which sheds light on cold-start recommendations.

We then propose a novel method for Explainable Recommendation based on Attribute Networks (\emph{eRAN}). In eRAN, items in an attribute network is represented by low-dimensional embedding vectors through a deep autoencoder to account for the nonlinearity and higher-order proximity in the network. Meanwhile, with users mapped to embedding vectors, an attention mechanism in adopted in eRAN to model user preferences toward different attributes. Then, a personalized item representation can be constructed by taking a weighted average from the node embeddings of each attribute network. Under the assumption that a user's co-purchased items should be more similar than others, eRAN optimizes towards the contrastive similarity to derive personalized similarity between item pairs, which is different from traditional item-to-item methods that directly factorize the item ratings through aggregate similarity~\cite{kabbur2013fism}. With the learned parameters, recommendation score of each item based on the users' embeddings and personalized item representations can be obtained and the most similar items in the lens of individual users would be recommended. Meanwhile, eRAN can provide fine-grained explanations on users' desired attributes for a particular recommended item. 

%{\color{red} Guannan, please reorganize the words in red and complete the above paragraph for technical descriptions of eRAN. Please give clear logics to the specific techniques adopted. What is the position in eRAN? for what? Also, do remember input the citations!}

%{\color{red}Furthermore, the personalized item representations can be constructed with the node embeddings in all the attribute networks. In spirit to item-to-item methods, we can also employ the neighborhood item representations to compute the individual-oriented similarity, aiming at minimizing the personalized similarity between the item pairs.
%In addition, non-paired items are also incorporated as negative samples, and we can further construct the ranking loss for the item relationships according to Bayesian Personalized Ranking (BPR)~\cite{rendle2009bpr}. With the learned parameters, we can score the items based on the derived users' embeddings and their neighborhood item representations, and meanwhile provide explanations on the desired attributes for a similar recommended item, which finally gives a method called {\bf E}xplainable {\bf R}ecommendation based on {\bf A}ttribute {\bf N}etworks ({\texttt{eRAN}}).}

We conduct experiments on three real-world datasets including \emph{movies}, \emph{books}, and \emph{music} to validate the effectiveness of the proposed methods, where the attributes can directly influence users' experiences towards these items. Experimental results demonstrate the superiority of the proposed methods to the state-of-the-art recommendation methods in terms of accuracy. Also, considering the incorporated auxiliary item attributes, the method is capable of coping with cold-start items when the attributes are given, and we further implement experiments to show that our method can better predict which users would be interested in these new items. Last but not least, we showcase a scenario how the attention weights and user embeddings inferred from the model can guide the explanations for recommendations.

% why SDNE?
%We need to capture the relationships between each co-purchased pair of items. Therefore, 

%Also, in most real-world e-commerce recommendation scenarios, the settings are generally implicit recommendation, in which a ranking-based for a pairwise comparison is needed. Thus, we adopt the BPR framework to construct the loss for the training.

\section{Attribute Network}

\begin{figure}[t!]
	\centering
	\includegraphics[width=0.48\textwidth]{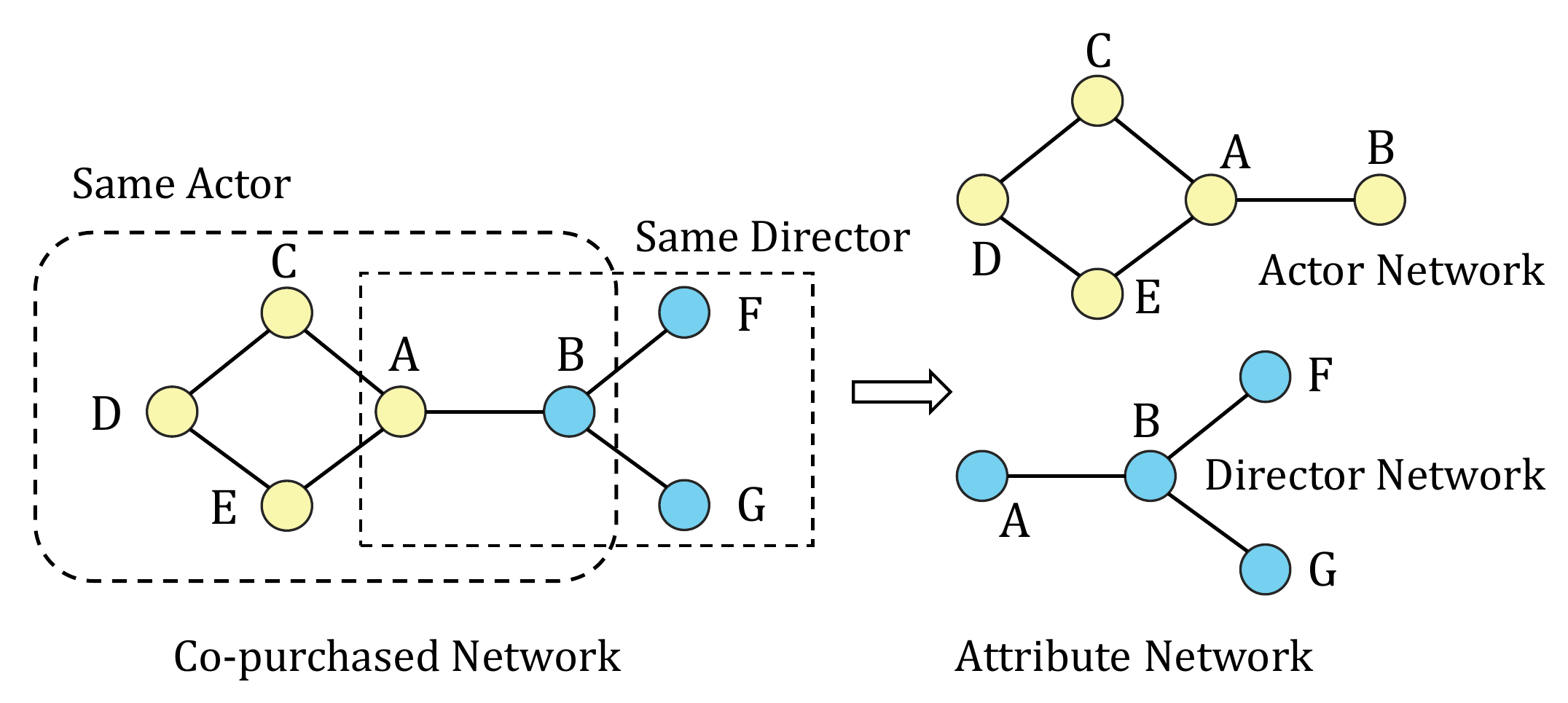}
	\caption{Toy example for attribute network.}
	\label{fig:toy_example}
\end{figure}
Items can be inter-connected with each other from various perspectives. For example, different items can appear in a same user's purchasing history, and the connections between the paired items can be driven by users' tastes and preferences. Also, connections can be established between items with similar attributes since they may both satisfy users' specific needs.
Such relationships can indeed be utilized to recommend items to individual users in a personalized manner with explanations on particular aspects.

% item to item.
One prominent strategy in constructing item relationships is to learn item similarity by factoring user's rating for an item as an aggregation of its similarity with previously rated items, or can be referred to as \emph{neighborhood items}~\cite{kabbur2013fism}. The relationships derived from item similarity can provide possible explanations for recommendations, \emph{e.g.}, ``\emph{item A is recommended because it is similar to the previously bought item B}''. Such relationships remain the same across users, \emph{i.e.}, every user would regard the similarity between each pair of items with no differences, which however, may not always hold and refrain the relationships to distinguish fine-grained user preferences. 

For example, the items are connected differently with regards to the graph structure as shown in Fig.~\ref{fig:toy_example}, where A has more connections with C, D, and E that have similar actors, while B interacts more with F and G due to their common directors. Two users X and Y may have watched both movies A and B, but the underlying reasons may be different. User X may only prefer the actors, while Y may be driven by the directors.
Thus, the relationships between A and B should be decomposed in accordance with users' preferences on particular aspects, \emph{i.e.}, the similarity between items should not be treated uniformly for every user.
In particular, for users such as X that favors the movies with specific actors, they may be more likely to accept the recommended item C or D, rather than F or G with the following explanations, ``\emph{We recommend \textbf{C} because it has similar \textbf{actors} with the watched movie \textbf{A}}''.

Obviously, item relationships can be decomposed by incorporating auxiliary information such as item attributes. With the attributes attached to each item, users' personalized preferences toward particular relationships can be explicitly explained. If the items with similar attributes are connected as shown in Fig.~\ref{fig:toy_example}, users' preferences can propagate along the connections driven by particular attributes. For example, user X's preferences toward actors can manifest in the local connections driven by \emph{actor}-network, while Y's favor for directors can be disclosed through the \emph{director}-network.

In addition, traditional content-based methods generally treat the item attributes independently and compute the similarity between items, which however, would lose higher-order relationships between them. Take the items again in Fig.~\ref{fig:toy_example}, item A and item D share no common actors, and the similarity based on the attribute of actors would be scored 0 in traditional sense. However, we can observe that A has a same actor with B, and B also has a same actress with C, which can indicate proximity between A and C.

Therefore, in order to capture users' preferences toward attributes and also account for and the higher-order relationships arise from particular attribute space, we propose a novel network structure, namely \emph{attribute network}.
% introduce the details of construting attribute network.
We firstly construct a \emph{co-purchased network} from the rating/purchasing history denoted by $\bm{G}=<\bm{V}, \bm{E}>$, where each item $i\in \bm{V}$ is regarded as a node, and ``also buy item $j$ with $i$'' is termed as a link between nodes $i$ and $j$, $e_{ij} \in \bm{E}$. In particular, item $i$ has a $K$-dimensional attribute vector $\mathbf{g}_i \in \mathbb{R}^K$, and for each type of attribute $k$, we only reserve the edges in $\bm{G}$ that share the same attribute values between the pairwise nodes, with the subset of links being $\bm{E}^k = \{e_{ij}^k | g_{ik}=g_{jk} ~\&\&~ e_{ij}\in \bm{E}\} \subseteq \bm{E}$, which gives an induced subgraph of $\bm{G}$, \emph{i.e.}, $k$-attribute network $\bm{G}^k$.

We assume that the reason why items are co-purchased can be attributed to one or several attributes. This assumption is indeed more applicable for experience products such as movies, music, \emph{etc.}, where users would show stable personalized preferences toward the attributes and the attributes are deemed to influence their experiences for the products greatly.
Given the $K$ attribute networks derived from the induced subgraphs of co-purchase networks, users' personalized preferences toward items can be decomposed as multi-attribute item relationships, and meanwhile the explanations for recommendations can be derived with regards to both item-based and content-based methods. 

\section{Methodology}
\begin{figure*}[t!]
	\centering
	\includegraphics[width=0.99\textwidth]{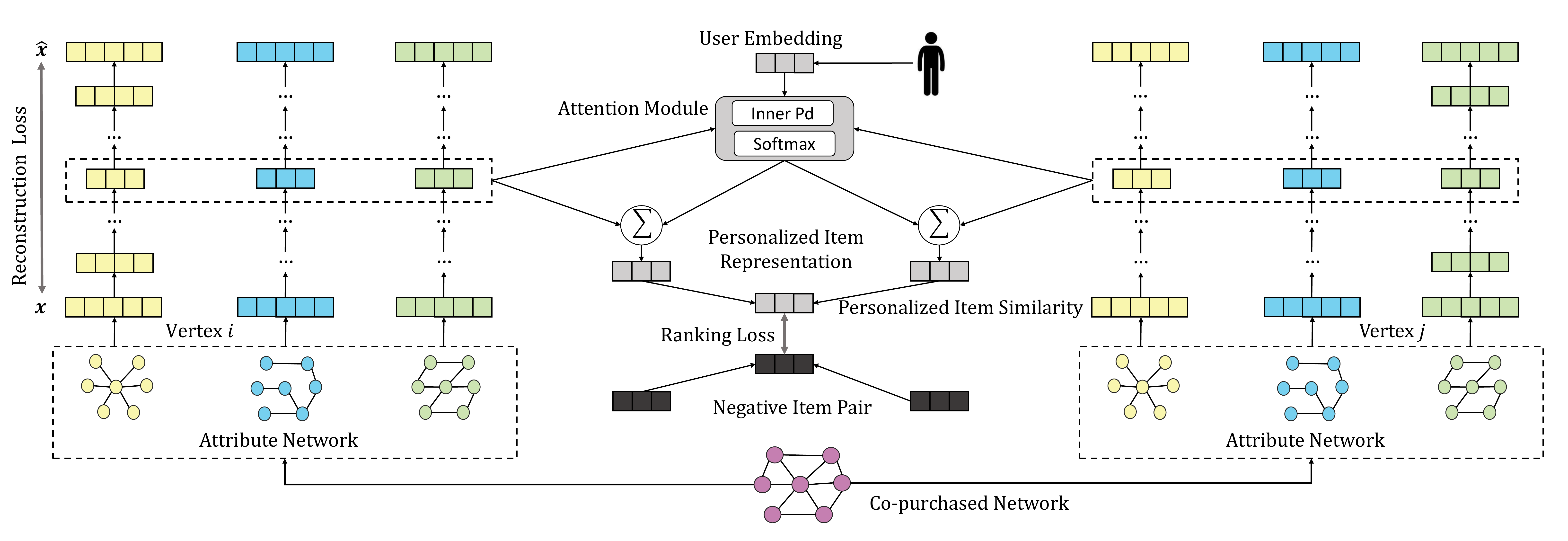}
	\caption{The architecture of eRAN.}
	\label{fig:framework}
\end{figure*}

The overview of the modeling framework is shown in Fig.~\ref{fig:framework}. Firstly each item in the attribute network is represented by K-attribute embedding vectors with deep autoencoders. Then, users are mapped to an embedding layer and the personalized preferences toward attributes are captured by an attention mechanism, so as to compute the personalized item similarity. Finally a personalized ranking loss is constructed to account for the individual item pairs and also apply negative sampling for non-paired items.

%Specifically, we first represent each item through their attribute network structure, how they connect to other items with similar attributes and how to represent the higher-order proximity from the attribute networks in particular. Moreover, considering the goal for recommendation, we also need to link users' preferences within the attribute network and further model users' preferences toward different perspectives of item relationships.

\subsection{Embedding Attribute Networks}
\label{sec:coldstartsec}
% why should embedding the attributes. the advantage of embedding the attribtue networks, rather than direclty embedding the items.
Attribute network provides a new perspective in probing users' preferences by mapping items to manifold space with different types of connections, which would overcome the limitations when handling raw item attributes in Euclidean space. Therefore, it is naturally appealing to firstly map items to low-dimensional vectors to encode the items in attribute space. For each node $v$ in attribute network $\bm{G}^k$, we can learn a mapping function $f^k(\cdot)$, to obtain a $d$-dimensional vector for the node, \emph{i.e.}, $\mathbf{h}_v^k \in \mathbb{R}^d$, where $d \ll |V|$.

As discussed previously, higher-order proximity in the attribute network can disclose the item similarity more accurately, and meanwhile users' preferences toward a specific attribute would propagate along the network path.
For example, in Fig.~\ref{fig:toy_example}, node A and node C share a common neighbor B, and they would be regarded as similar in the connections when second-order similarity is taken into account. Specifically, second-order proximity can be defined as the similarity of the neighborhood structure between a pair of nodes, thus we can represent a node by its adjacency vector $\mathbf{s}_v^k \in \bm{S}^k$. $\bm{S}^k$ is the adjacency matrix constructed from the $k$-attribute network, and the entry $s_{ij} = 1$ when there exists a link between node $i$ and $j$.

Considering the high nonlinearity of the network structure, we propose to represent the adjacency vector of each node in the attribute network via a deep autoencoder.
% revisit the technical details of AE.
Deep autoencoder is a typical deep learning model to handle nonlinearity, which generally consists of two parts including encoder and decoder, with both containing multi-layer nonlinear functions. The feedforward process of the encoder maps the input data $\mathbf{x}$ to the representation space as follows,

\begin{equation}
\begin{split}
&\mathbf{y}^{(1)} = \sigma(W^{(1)} \mathbf{x}+b^{(1)}) \\
&\mathbf{y}^{(t)} = \sigma(W^{(t)} \mathbf{y}^{(t-1)} + b^{(t)}),~~t=2,...,T,
\end{split}
\label{eq:ae}
\end{equation}
where $T$ represents the number of layers for encoder, and $W^{(t)}, b^{t}$ denotes corresponding parameters of the layer. Particularly, $\mathbf{y}^{(t)}$ can be regarded as the hidden representation for $\mathbf{x}$ when $t=T$. In similar to the encoder part, decoder also applies several non-linear functions to mapping the representation vectors to the reconstruction space, and obtain the reconstructed output $\hat{\mathbf{x}}$. By minimizing the reconstruction error between the input and the output, we can derive the representation, and the loss function can be formulated as follows.

\begin{equation}
\mathcal{L} = \sum_{i=1}^{N}||\mathbf{x}_i - \hat{\mathbf{x}_i} ||^2_2.
\end{equation}

We treat the adjacency vector $\mathbf{s}_v$ of the node $v$ in the $k$-attribute as input and feed it into the autoencoder to derive the hidden representations for the item in the $k$-attribute network.

% accounting for zeros.
However, the network may be sparse and the adjacent matrix would be filled with many zeros. Thus, traditional autoencoder may be more likely to reconstruct zeros and hence cannot capture the local connectivity of the network structure. In order to tackle this issue, we impose a larger penalty on the reconstruction error of the non-zero elements~\cite{wang2016structural} by incorporating a regularizer $\mathbf{b}=\{b_{ij}\}$. When $s_{ij}=0$, $b_{ij}$ is set to be a small value $\beta$, and the loss can be revised as follows.

\begin{equation}
\mathcal{L}_{net} = \sum_{i=1}^{N}||(\mathbf{x}_i - \hat{\mathbf{x}_i}) \odot \mathbf{b}_i||^2_2 \label{netloss}
\end{equation}
where $\odot$ means the Hadamard product.

%reconstruction loss with specific concerns for avoiding constructing too many zeros. 

% handling newly-released items, with known attributes. The emebdding vectors can be constructed by forward in the deep auto-encoders.
It is worth addressing that this attribute-network representation framework is capable of handling cold-start items with attributes given. Though a newly released item has no prior co-purchase records, the attributes provide clues to connect it with existing attribute networks. Specifically, we can regard the item as a new node $v^\prime$ with attribute vector $\mathbf{g}_{v^\prime}$. Then for the $k$-attribute network, the edges can be connected to those existing nodes that have the same attribute value with $g_{v^\prime}^k$. With the derived parameters of the autoencoder structure, we can further obtain the hidden representation $\mathbf{h}_{v^\prime}^k$ of the new item.

\subsection{Personalized Item Similarity Based on Attribute Networks}
Item-to-item CF is a typical method by employing the \emph{neighborhood items} to compute the recommendation score for an item. In this approach, the item similarity is computed by the inner dot between the latent factors of the item pairs, which is generally in the following form according to~\cite{kabbur2013fism}.

\begin{equation}
sim(i,j) = \mathbf{p}_j\mathbf{q}_i^\top,
\end{equation}
where $\mathbf{p}_j$ and $\mathbf{q}_i$ denote the latent factors of items respectively. The similarity between item $i$ and $j$ obtained through the inner dot is indeed uniform across distinct users, which remain to be a major limitation for these methods. Therefore, it is desirable to take an individual view to account for the item similarity.

With the derived representations from attribute networks, we can replace the item latent factors with the node embeddings to devise a neural model for item-based recommendation. Moreover, we can decompose the item-to-item similarity by taking users' preferences toward each field of attribute into account. Each user $u$ can firstly be mapped to an embedding vector $\mathbf{z}_u$, and then the user's preferences toward a particular attribute of an item can be captured through an attention mechanism. Attention mechanisms are generally introduced in NLP, computer vision, and recommender systems to track the attractions of different components.
Specifically, in scoring the attention weights of user $u$ for item $i$ on a particular attribute $k$, we simply take the inner dot between the user embedding vectors and the node embeddings $\mathbf{h}_i^k$  from the deep autoencoder of the $k$-attribute as follows,
\begin{equation}
a_{u,i}^k = {\mathbf{h}_i^k} \mathbf{z}_u ^\top
\label{eq:attention}
\end{equation} 

We can further apply softmax to normalize user's attention scores for an item on each attribute $k$,
\begin{equation}
\tilde{a}_{u,i}^k = \frac{\exp(a_{u,i}^k)}{\sum_{k^\prime=1}^{K}\exp(a_{u,i}^{k^\prime})}
\label{eq:attrnorm}
\end{equation}

The attention weights can be explained as the extent to which the user desire for a particular attribute of an item, and thus can be exploited to provide explanations for the attribute-aware recommendation.

Then, we can proceed to derive each individual user' affinity for an item through the node embeddings from all the attribute network, which can be viewed as a weighted average of the node embeddings from each type of attribute network.
\begin{equation}
\mathbf{h}_{u,i} = \sum_{k=1}^{K} \tilde{a}_{u,i}^k \mathbf{h}_{i}^k
\end{equation}

Motivated by the general idea of item-to-item methods, we can employ the item representations to compute the personalized similarity in the neighborhood, which can be approximated by,

\begin{equation}
sim_u(i,j) = -||\mathbf{h}_{u,i} - \mathbf{h}_{u,j} ||^2_2 
\label{eq:simi},
\end{equation}

Different from prior item-based methods, we replace the inner dot with L2-norm distance metric to measure the item relationships with the embedding vectors. As proved in ~\cite{Hsieh2017Collaborative}, inner dot violates the triangle inequality, which may lead to suboptimal solution. Moreover, the personalized item similarity indeed decomposes the relationships towards attributes, which can provide fine-grained explanations for recommendations.

\subsection{Loss Function and Optimization}
Since we obtain the personalized similarity for each pair of items, we can use it to guide the learning of both user embeddings and item representations, as well as the attention weights on different attributes. An underlying assumption is that users would remain stable in their preferences for the items and therefore the neighborhood item tend to be similar in view of the users. Specifically, given a particular user $u$ and one of the purchased item $i \in \mathcal{R}_u^+$, it should be similar to the neighborhood items $j \in \mathcal{R}_u^+$. Thus, the representations can be learned by maximizing the aggregate personalized similarity, and can be written as a loss function by taking a negative value based on Equation~\eqref{eq:simi},

\begin{equation}
\mathcal{L} = \sum_{u \in \mathcal{U}} \sum_{i,j \in \mathcal{R}_u^+} \log\sigma(||\mathbf{h}_{u,i} - \mathbf{h}_{u,j} ||^2_2),
\end{equation}
where $\mathcal{U}$ represents all the users. However, this loss function is likely to get trapped to a trivial solution when all the items are approximated by the same representation. Thus, similarly to the optimization techniques proposed in BPR \cite{rendle2009bpr} that assume users would prefer items they have bought than those that they have not, we also introduce a negative sampling strategy to avoid the issues. 

Specifically, given a user $u$, we can sample an item $n \notin \mathcal{R}_u^+$ as a negative sample. Then for each item $i$ and a co-purchased item $j$ by $u$, it is natural that the similarity should be higher than that between the non-paired items $i$ and $n$, which would satisfy the following inequality.
\begin{equation}
sim_u(i,j) >_u sim_u(i,n).
\end{equation}

Then the loss function with negative sampling can be revised as,
\begin{equation}
\mathcal{L}_{rank} = \sum_{u \in \mathcal{U}} \sum_{i,j \in \mathcal{R}_u^+} \log \sigma(||\mathbf{h}_{u,i} - \mathbf{h}_{u,j} ||^2_2 - ||\mathbf{h}_{u,i} - \mathbf{h}_{u,n} ||^2_2).
\label{rankloss}
\end{equation}

To preserve the attribute network structures and learn personalized item presentations tailored for recommendation, we combine the loss functions in Equation \eqref{netloss} and Equation \eqref{rankloss} with a weighting parameter $\alpha$ to jointly minimizes the following objective function:

\begin{equation}
\mathcal{L}_{rec} = \mathcal{L}_{net} + \alpha \mathcal{L}_{rank} \label{lossall}.
\end{equation}

We adopt Adaptive Moment Estimation (Adam) to optimize the objective function in Equation \eqref{lossall}. In each iteration, we sample a mini-batch of users and item pairs with its corresponding $k$ adjacency matrix to update the parameters.

\subsection{Recommendation Score}
Similar to traditional item-to-item CF method, when evaluating the recommendation score of user $u$ for item $i$ given the learned representations, we need to revisit the relationships between item $i$ and each item $j$ that has ever been purchased by $u$, which can be approximated by,

\begin{equation}
\hat{r}_{u,i} = \sum_{j\in \mathcal{R}_u^+\backslash\{i\}} -||\mathbf{h}_{u,i} - \mathbf{h}_{u,j} ||^2_2,
\end{equation}
%特别地，对于冷启动产品，容易被主导，所以可以被计算为：
%当推荐的时候，simply选择最小的topN个rui作为推荐列表即可 // 后续再说不同之处也可
where $\mathcal{R}_u^+\backslash\{i\}$ represents the set of items that have been rated by the user $u$ except for $i$, \emph{i.e.}, the neighborhood of item $i$. 
%And $C$ is a constant large enough to maintain the defination of the recommendation score. 

In particular, for a new item $x$, since it does not have any prior co-purchase records, we can only connect it to the existing nodes in the attribute networks. According to the learned parameters and representations, we can also construct the individual representation $\mathbf{h}_{u, x}$ for user $u$. However, item $x$ never appears together with any of neighborhood items, thus we relax the restriction and derive the recommendation score with regards to the minimum similarity with the neighborhood items.
%it's difficult to be similar with each item in $\mathcal{R}_u^+\backslash\{i\}$, thus the recommendation score can be evaluated by,
\begin{equation}
\hat{r}_{u,x} = \min_{j\in \mathcal{R}_u^+} -||\mathbf{h}_{u,x} - \mathbf{h}_{u,j} ||^2_2
\label{eq:coldstartrec}
\end{equation}

When generating recommendations for $u$, we simply need to rank candidate items according to the recommendation score and select the ones with highest scores as recommended items.
%	may need to discuss how we can present explanations from this recommendation framework?
In this recommendation framework, we can easily interpret the recommendations with both the personalized item similarity and the users' attention weights on attributes of each item. Specifically, when the item $i$ is to be recommended to user $u$, we can obtain the attention weights $\tilde{a}^k_{u,i}$ according to Equation~\eqref{eq:attention} and \eqref{eq:attrnorm}, to identify which attributes of $i$ that attract the user; meanwhile, we can also position the item $j$ in the neighborhood that are most similar to $i$. Therefore, we can recommend $i$ to $u$ with the following explanations:
``\emph{We recommend $i$ because it is similar to $j$ on the attribute $k$}.

\section{Experimental Setup}
We validate the effectiveness of the proposed methods on three real-world datasets. Eight state-of-the-art (SOTA) baseline methods are included for a thorough comparative study.
\subsection{Data Sets} 
% We demonstrate the effectiveness of our model on three real-world datasets, with data statistics listed in Table~\ref{tab:data}. 

We first briefly introduce the datasets used in our experiments, with the statistics listed in Table~\ref{tab:data}.

\textbf{Kaggle-Movie}: This dataset is extracted from the Kaggle\footnote{https://www.kaggle.com/rounakbanik/the-movies-dataset/} Challenge Dataset. We use the directors, genres and top five actors as attributes.

% \textbf{Goodreads-Potery}: This dataset is collected by ~\cite{wan2018item} from Good-reads\footnote{https://www.goodreads.com}, a popular online book review website. Several metadatas are used as features, including authors, number of pages, publication-year, and top three user-generated shelf names.
\textbf{Goodreads-Potery}: This dataset is collected by Wan et al.~\cite{wan2018item} from a popular online book review website named Good-reads\footnote{https://www.goodreads.com}. Several attributes are used including authors, number of pages, publication-year, and top three user-generated shelf names.

% \textbf{Amazon-Music}: Top-level product categories on Amazon\footnote{https://www.amazon.com} are constructed as a group of separate datasets by ~\cite{mcauley2015image,he2016ups}. In this paper, we choose the Music dataset and use both the top three genres and price as features.
\textbf{Amazon-Music}: Each top-level product category on the Amazon\footnote{https://www.amazon.com} are constructed as a separate datasets by McAuley et al.~\cite{mcauley2015image}. We choose the dataset constructed from the music category, and extract the top three genres as well as the price as attributes.

% In this paper, we remove items with missing values, convert ratings larger than 3 to positive feedbacks, and retain users whose history length larger than 5, 5, 3 for Kaggle-Movie, Goodreads-Potery and Amazon-Music respectively.
In this paper, we remove items with missing values, treat ratings larger than 3 as positive feedbacks, and retain users whose history length larger than 5, 5 and 3 for Kaggle-Movie, Goodreads-Potery and Amazon-Music respectively.

\begin{table}[]
	\centering
	\caption{Data statistics.}
	\vspace{-0.2cm}
	\label{tab:data}
	\begin{tabular}{lcccc}
		\toprule
		& \# users & \# items & \# actions & \# features \\ \hline
		Kaggle-Movie & 663 & 6850 & 61088 & 3 \\ 
		Goodreads-Potery & 39540 & 24052 & 449401 & 4 \\ 
		Amazon-Music & 11697 & 7100 & 65950 & 2 \\\bottomrule
	\end{tabular}
	\vspace{-0.3cm}
\end{table}

\subsection{Baseline Methods}
The following SOTA methods are applied as baselines in our experiments.

\textbf{NMF}~\cite{Paatero2010Positive}: NMF is a widely used collaborative filtering approach, which factorizes the interaction binary matrix.

\textbf{BPR-MF}~\cite{rendle2009bpr}: BPR-MF is a well-known top-N recommendation method to cope with implicit matrix, which uses the Bayesian personalized ranking optimization criterion.

\textbf{FM}~\cite{rendle2010factorization}: FM is a successful feature-based recommendation method, which is effective on sparse data.

\textbf{DeepFM}~\cite{guo2017deepfm}: DeepFM is a deep variant of FM which imposes a factorization machines as "wide" module to extract shallow feature interactions.

\textbf{PNN}~\cite{qu2016product}: PNN is another deep variant of FM which introduces a product layer after embedding
layer to capture high-order feature interactions.

\textbf{AFM}~\cite{xiao2017attentional}: AFM extends FM by using attention mechanism to distinguish the different importance of
second-order combinatorial features. 

\textbf{SVDFeature}~\cite{chen2012svdfeature}: SVDFeature is an effective toolkit for feature-based matrix factorization. 

\textbf{FISM}~\cite{kabbur2013fism}: FISM is a state-of-the-art item-based CF method which learns global item similarities from user-item interactions.

\textbf{eRAN-L1}: eRAN-L1 is a submodel which only optimizes the ranking loss.

\textbf{eRAN-L2}: eRAN-L2 is another submodel which only optimizes the reconstruction loss with $\alpha=0$. In this submodel, we fix user embeddings to 1.0 during training.

\subsection{Parameter Settings}
% For our method, we set the mini-batch size, the learning rate of the Adam, the hyper-parameters of $\alpha$ and $\beta$ to be 2000, 0.001, 1500 and 0.2 respectively. The structure of autoencoder is the same for all the three datasets, namely \#items-1024-256-32. For other baseline methods, we apply default parameters except for the embedding size, which is fixed to be 32 for all the methods.
For our method, we set the mini-batch size, the learning rate of the Adam, the hyper-parameters of $\alpha$ and $\beta$ to be 2000, 0.001, 1500 and 0.2 respectively. We keep the same structure of autoencoder with varying datasets. Specifically, the dimensions of hidden states are 1024, 256 and 32 for $\mathbf{y}^{(2)},\mathbf{y}^{(3)}$ and $\mathbf{y}^{(4)}$ respectively according to Equation~\eqref{eq:ae}. As for the baseline methods, we apply default parameters except for the embedding size, which is fixed to be 32 for all the methods.

%\subsection{Tasks and Evaluation Measures}
% We first validate the proposed model in comparison with several state-of-the-art methods. Then, by conducting a new item recommendation task, we evaluate the performance of the model on cold start problems. We also explore the learned user embeddings and attenions from both quantitative and qualitative perspectives to explain the recommendation. Finally, we seek the influences of some important parameters. For new item recommendation task, we apply Recall@K as measure while Precision@K and NDCG@K for other tasks.

%We first conduct a comparative study to validate the superiority of our model, where eight baseline methods are included. Then, we design a new item recommendation task to evaluate the performance of our model when facing cold start issue. We also explore the learned user embeddings and attention weights from both quantitative and qualitative perspectives to show our model is capable of explaining the recommendation. Finally, we study the impacts of some important parameters. For new item recommendation task, we apply Recall@K as measure while Precision@K and NDCG@K for other tasks given top-K recommendations.

\section{Experimental Results}
\begin{table*}[t!]
	\centering
	\scriptsize
	\caption{Precision@K of the three datasets.}
	\label{tab:rec_precision}
	\resizebox{0.85\textwidth}{!}{
		\begin{tabular}{@{}llccccccccc@{}}
			\toprule
			\multirow{2}{*}{Method} & \multicolumn{3}{c}{{\it Kaggle-Movie}} & \multicolumn{3}{c}{{\it Goodreads-Potery}} & \multicolumn{3}{c}{{\it Amazon-Music}} \\
			\cmidrule{2-10}
			&P@5 & P@10 & P@15 & P@5 & P@10 & P@15 & P@5 & P@10 & P@15 \\ \midrule
			NMF & 0.1201 & 0.0746 & 0.0548 & 0.1386 & 0.0779 & 0.0525 & 0.0902 & 0.0602 & 0.0454 \\
			BPR-MF & 0.1210 & 0.0742 & 0.0547 & 0.1412 & 0.0801 & 0.0565 & 0.0806 & 0.0516 & 0.0392  \\
			FISM & 0.1217 & 0.0736 & 0.0536 & 0.1528 & 0.0827 & 0.0576 & \underline{0.0951} & 0.0636 & 0.0465 \\
			\cmidrule{1-10}
			FM & 0.1168 & 0.0725 & 0.0531 & 0.1524  & 0.0844 & 0.0578 & 0.0844 & 0.0530 & 0.0386\\
			DeepFM & 0.1183 & 0.0726 & 0.0542 & 0.1540 & \underline{0.0849} & 0.0590 & 0.0874 & 0.0559 & 0.0419 \\
			PNN & 0.1195 & 0.0719 & 0.0537 & \underline{0.1557} & 0.0842 & \underline{0.0590} & 0.0875 & 0.0571 & 0.0428 \\
			AFM & 0.1154 & 0.0721 & 0.0528 & 0.1376 & 0.0783 & 0.0550 & 0.0739 & 0.0497 & 0.0387 \\
			SVDFeature & \underline{0.1219} & \underline{0.0751} & \underline{0.0556}  & 0.1547 & 0.0848 & 0.0588 & 0.0943 & \underline{0.0637} & \underline{0.0480}\\
			\cmidrule{1-10}
			eRAN-L1 & 0.1161 & 0.0733 & 0.0532 & 0.1485 & 0.0836 & 0.0572 & 0.0707 & 0.0470 & 0.0369\\
			eRAN-L2 & 0.0237 & 0.0190 & 0.0161 & 0.0305 & 0.0218 & 0.0163 & 0.0541 & 0.0385 & 0.0305\\
			eRAN & \textbf{0.1289} & \textbf{0.0789} & \textbf{0.0570} & \textbf{0.1626} & \textbf{0.0875} & \textbf{0.0604} & \textbf{0.1104} & \textbf{0.0691} & \textbf{0.0508} \\ \bottomrule
	\end{tabular}}
\end{table*}

\begin{table*}[t!]
	\centering
	\scriptsize
	\caption{nDCG@K of the three datasets.}
	\label{tab:rec_ndcg}
	\resizebox{0.85\textwidth}{!}{
		\begin{tabular}{@{}llccccccccc@{}}
			\toprule
			\multirow{2}{*}{Method} & \multicolumn{3}{c}{{\it Kaggle-Movie}} & \multicolumn{3}{c}{{\it Goodreads-Potery}} & \multicolumn{3}{c}{{\it Amazon-Music}} \\
			\cmidrule{2-10}
			&n@5 & n@10 & n@15 & n@5 & n@10 & n@15 & n@5 & n@10 & n@15 \\ \midrule
			NMF & 0.4451 & 0.4986 & \underline{0.5213} & 0.5837 & 0.5988 & 0.6149 & 0.3280 & 0.3697 & 0.3941 \\
			BPR-MF & 0.4554 & 0.5002  & 0.5206 & 0.6044 & 0.6256 & 0.6379 & 0.2807 & 0.3173 & 0.3362 \\
			FISM & \underline{0.4593} & \underline{0.5017} & 0.5195 & \underline{0.6579} & \underline{0.6783} & \underline{0.6923} & \underline{0.3672} & \underline{0.4075} & \underline{0.4253} \\
			\cmidrule{1-10}
			FM & 0.3639 & 0.4200 & 0.4419  & 0.6222 & 0.6508 & 0.6595 & 0.2971 & 0.3324 & 0.3454\\
			DeepFM & 0.3782 & 0.4267 & 0.4503 & 0.6370 & 0.6627 & 0.6731 & 0.2969 & 0.3363 & 0.3547\\
			PNN & 0.3822 & 0.4341 & 0.4586 & 0.6527 & 0.6747 & 0.6827 & 0.3015 & 0.3431 & 0.3618 \\
			AFM & 0.3727 & 0.4289 & 0.4527 & 0.5552 & 0.5857 & 0.5968 & 0.2437 & 0.2867 & 0.3079 \\
			SVDFeature & 0.4272 & 0.4755 & 0.4964  & 0.6464 & 0.6716 & 0.6873 & 0.3370 & 0.3888  & 0.4122 \\
			\cmidrule{1-10}
			eRAN-L1 & 0.3709 & 0.4312 & 0.4491 & 0.6174 & 0.6581 & 0.6693 & 0.2164 & 0.2622 & 0.2844\\
			eRAN-L2 & 0.0684 & 0.0916 & 0.1064 & 0.0852 & 0.1019 & 0.1467 & 0.1805 & 0.2173 & 0.2367\\
			eRAN & \textbf{0.4702} & \textbf{0.5167} & \textbf{0.5340} & \textbf{0.6858} & \textbf{0.7073} & \textbf{0.7154}  & \textbf{0.4026} & \textbf{0.4482} & \textbf{0.4671} \\ \bottomrule
	\end{tabular}}
\end{table*}

\subsection{Recommendation Accuracy}
We first conduct a comparative study to validate the superiority of our model to the introduced baseline methods in terms of recommendation accuracy.
In this task, we adopt the leave-one-out evaluation strategy, that is, for each user, we hold-out one purchased item as test set and the remaining is used for training. Since it is too time-consuming to rank all the items for every user during evaluation, we follow the experimental settings in~\cite{he2017neural} which randomly samples 100 negative items and rank the recommendation score among the 100 items. Given the top-K ranked items, we apply Precison@K and nDCG@K as evaluations measures. The comparative results of the three datasets are in Table~\ref{tab:rec_precision} and Table~\ref{tab:rec_ndcg}.

The proposed model is consistently better than all the baselines on the three datasets, while in contrast, the second best is relatively unstable, showing that our methods are more robust. In addition, we find that FISM outperformed other baselines in many cases in nDCG, while SVDFeature and PNN performed better in Precision. This results indicate that eRAN can not only accurately recognize the items that users really prefer, but also tend to rank them at top positions simultaneously.

Moreover, we find that most attribute-based methods perform well on {\it Goodreads-Potery} particularly. A possible reason is that the attribute \emph{user-generated shelf names} and \emph{authors} may have great influences on user preferences, and our method can effectively infer users' preferences toward attributes.
Also, it is notable that eRAN achieves the greatest improvement on {\it Amazon-Music} with the most sparse ratings among the three datasets. It might be due to the attribute network simultaneously model the first-order relationship and the high-order relationship from attribute space, which can handle the data sparsity.

It is also notable that AFM doesn't perform well on the three datasets, even worse than FM, which may be due to the fact that it is unable to learn effective attention weights in feature interaction space when features are scarce. While on the contrary, eRAN can leverage attention mechanism to model user's fine-grained preferences in the attribute space.

\begin{figure}[t!]
	\centering
	\includegraphics[width=0.4\textwidth]{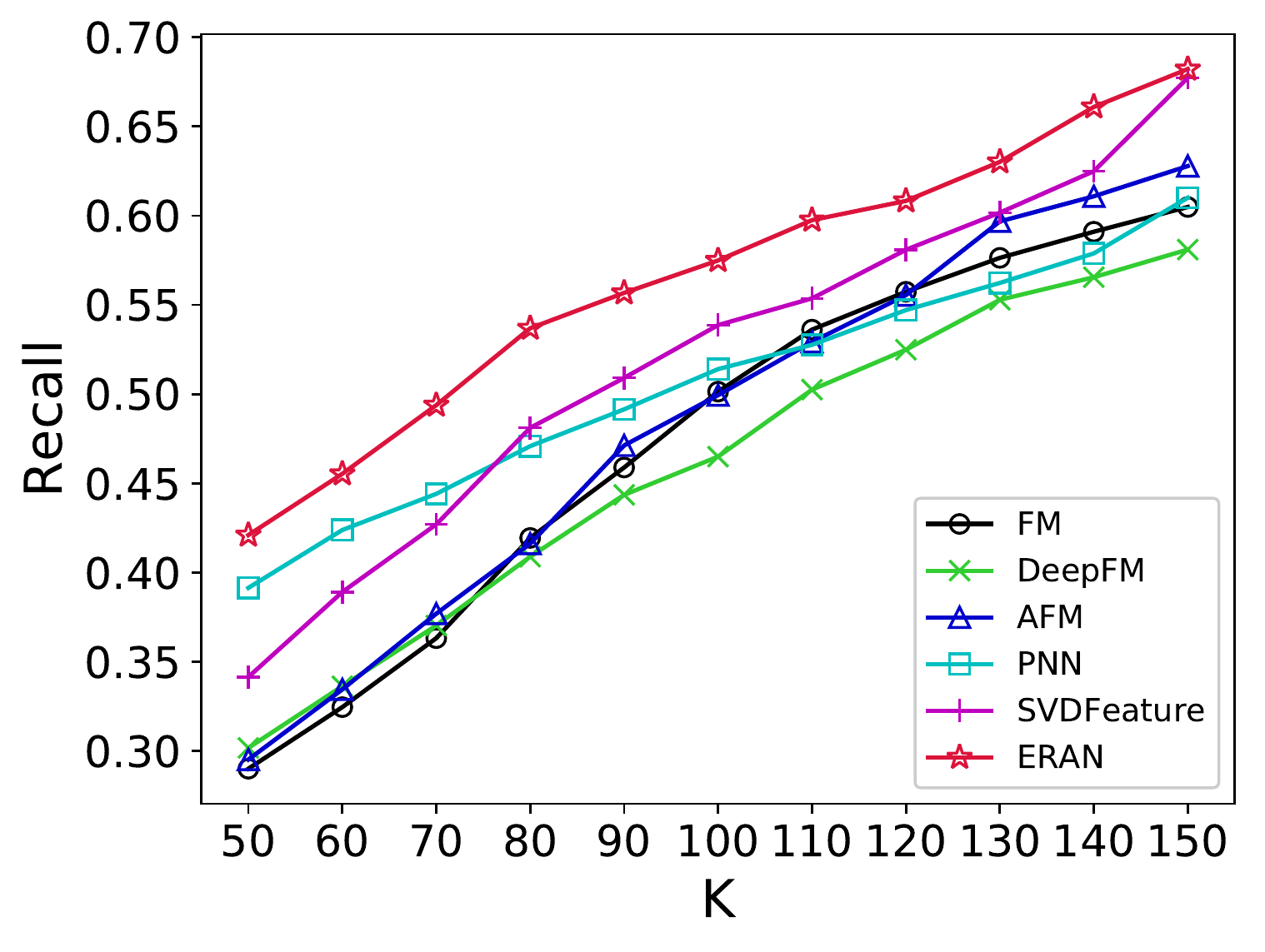}
	\caption{Prediction results for cold-start items.}
	\label{fig:cold_start}
\end{figure}

Considering the two variants of eRAN, the submodel eRAN-L2 can be seen as a kind of network embedding method which lacks optimization tailored for recommendation, which achieves the worst performances. Meanwhile, eRNA outperforms eRAN-L1 significantly, which validates the effectiveness of leveraging attribute information in improving performances.

\subsection{Cold Start Item Recommendation}
In this task, we evaluate the effectiveness of our model in handling cold start items with attributes given. To simulate the scenarios for cold-start items, we randomly hold out 40 items and regard them as new items with no purchasing records. We can treat the item as a new node and connect it with existing attribute networks according to Section~\ref{sec:coldstartsec}. Afterwards, we can obtain the adjacency matrix of the new node in each attribute network and feed into the deep autoencoder to derive respective node embeddings.

Then, for each `new' item, we can rank all the users according to the recommendation score in Equation~\eqref{eq:coldstartrec} to predict which users are most likely to purchase the items. We can use the measure \emph{Recall} to evaluate the effectiveness of prediction, \emph{i.e.}, how many users are accurately predicted to buy the new item among all the users that have purchased in reality.
The methods NMF, BPR, and FISM cannot be applied for this setting because the new items would not appear in the rating matrix. Thus, we only implement this experiment on the attribute-based recommendation methods, in which we remove the corresponding new items in the training phase and evaluate the results on prediction with the same setting of our method.

The results on {\it Kaggle-Movie} are illustrated in Figure~\ref{fig:cold_start}. As we can see that our methods consistently outperform other attribute-based recommendation methods.
Among the baselines, FM, DeepFM and AFM achieve similar performances in this experiment, PNN performs the second best when $K$ is small, while SVDFeature shows competitive results when $K$ is large. The results show the superiority of the proposed method in coping with cold-start items, which also illustrate that eRAN well capture fine-grained user preferences toward attributes.

%As the results show, our method outperforms other feature-based methods in all recommendation sizes. We point out that it may benefit from the attention mechanism. Baselines assign the same feature weights for all users, which implies that preferences on particular attributes are the same for all users. This is unreasonable for the reason analyzed in the above sections. Some users may prefer a certain aspect such as actors while others may not. A cold start movie with a movie star may be more attractive for the users who prefer actors more when watching a movie. We apply the attention mechanism to learn personalized feature weights for every user so that it can capture such fine-grained user preferences and obtain the best performance. 

\subsection{Explanation and Visualization}

One of the pervasive advantages our model is that we can obtain insights into the underlying reasons for recommendation. Thus, we explore the learned user embeddings and attention weights on attributes from both quantitative and qualitative perspectives to explain the recommendation. Take the {\it Kaggle-Movie} dataset as an example, for each user, we regard the users' average attention scores for all the items that they have interacted with as the general description of their preferences. Correspondingly, each user in the movie dataset can be described with attentions scores on \emph{actor}, \emph{director}, and \emph{genre}.
Larger attention score on an attribute means the user may prefer the corresponding aspect more.

\begin{table*}[!t]
	\centering
	%\scriptsize
	\LARGE
	\caption{Two comparative case studies for explainable recommendation.}
	\label{tab:case_study}
	\resizebox{0.98\textwidth}{!}{
		\begin{tabular}{@{}llccccccccc@{}}
			\toprule
			Movie & User & Actor Attention Score & Director Attention Score & Most Similar Movies & Explanation \\ \midrule
			\multirow{2}{*}{Fear and Loathing in Las Vegas} & 255 & \textbf{0.6006} & 0.3021 & Edward Scissorhands, A Nightmare on Elm Stree & The same actor \textbf{Johnny Depp} \\ 
			& 639 & 0.3404 & \textbf{0.5159} & The Meaning of Life, Monty Python and the Holy Grail & The same director \textbf{Terry Gilliam} \\ \midrule
			\multirow{2}{*}{Pulp Fiction} & 467 & \textbf{0.4941} & 0.2874 & Django Unchained, Jurassic Park & The same actor \textbf{Samuel L. Jackson} \\
			& 129 & 0.3091 & \textbf{0.4708} & Kill Bill, Reservoir Dogs & The same director \textbf{Quentin Tarantino} \\ \bottomrule
	\end{tabular}}
\end{table*}

\begin{figure*}[t!]
	\centering
	\begin{subfigure}[b]{0.32\textwidth}
		\includegraphics[width=\textwidth]{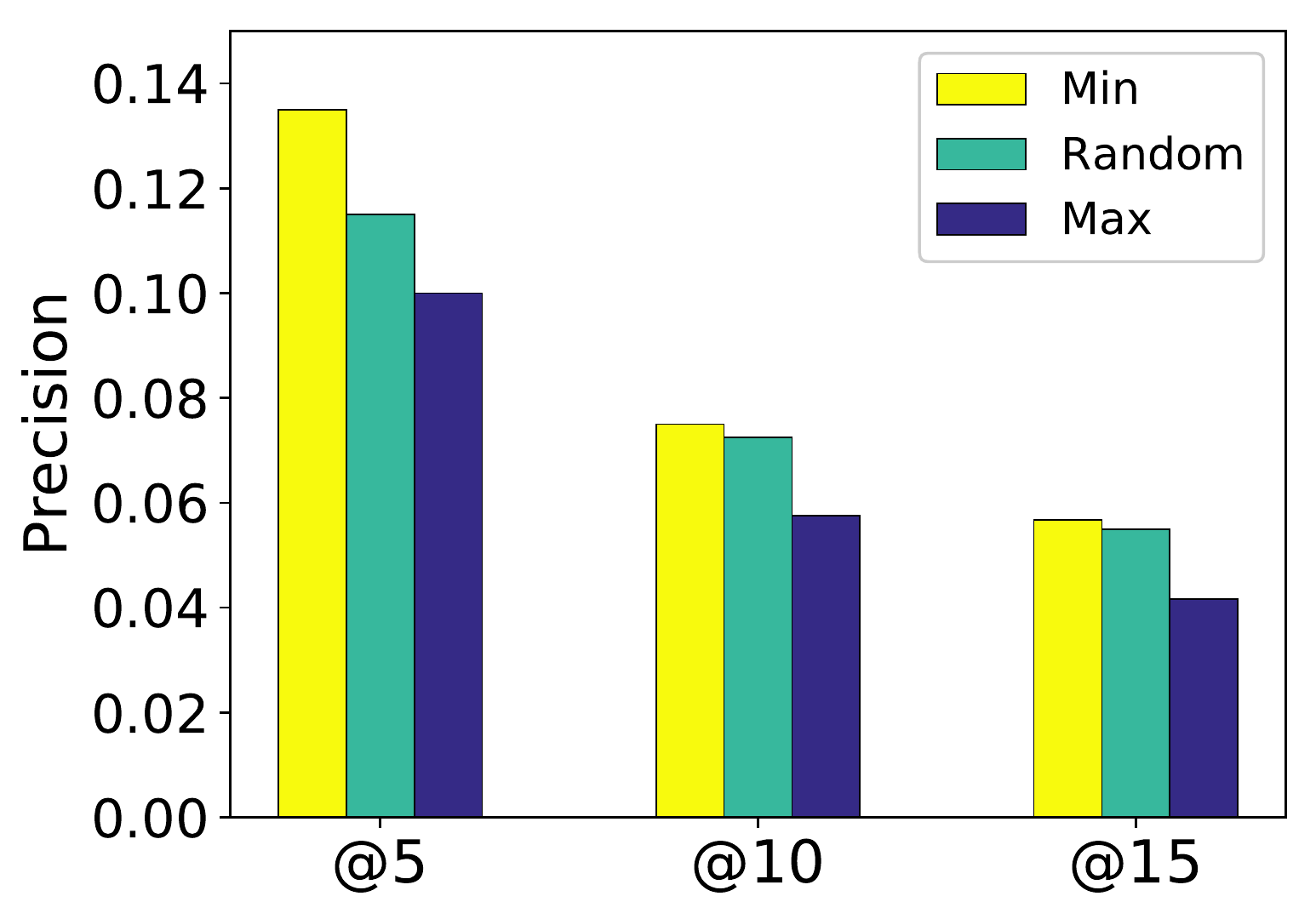}
		\caption{Precision}
		\label{fig:ex_precision}
	\end{subfigure}
	%	\hspace{0.04\textwidth}
	\begin{subfigure}[b]{0.32\textwidth}
		\includegraphics[width=\textwidth]{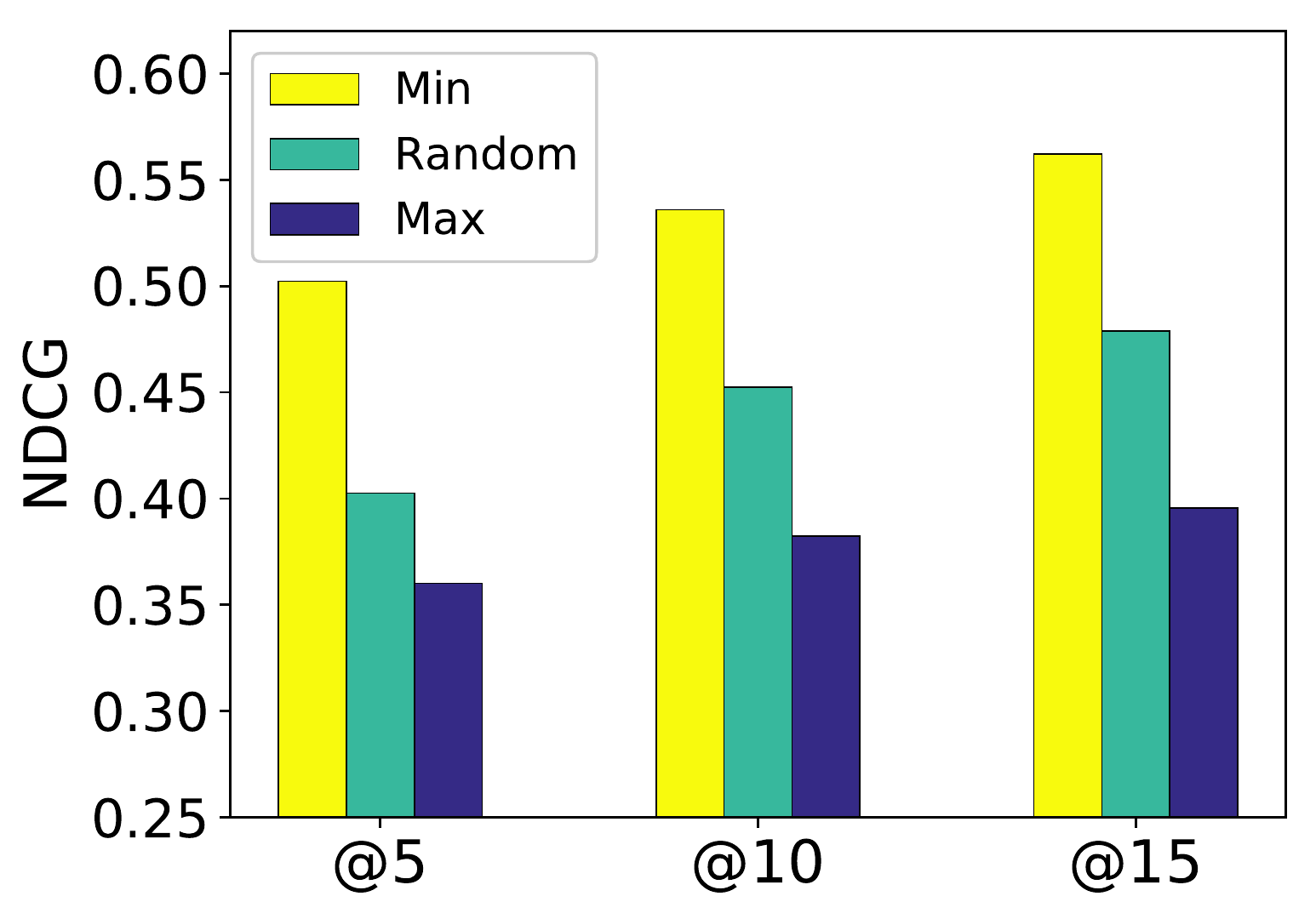}
		\caption{nDCG}
		\label{fig:ex_ndcg}
	\end{subfigure}
	%	\hspace{0.04\textwidth}
	\begin{subfigure}[b]{0.3\textwidth}
		\includegraphics[width=\textwidth]{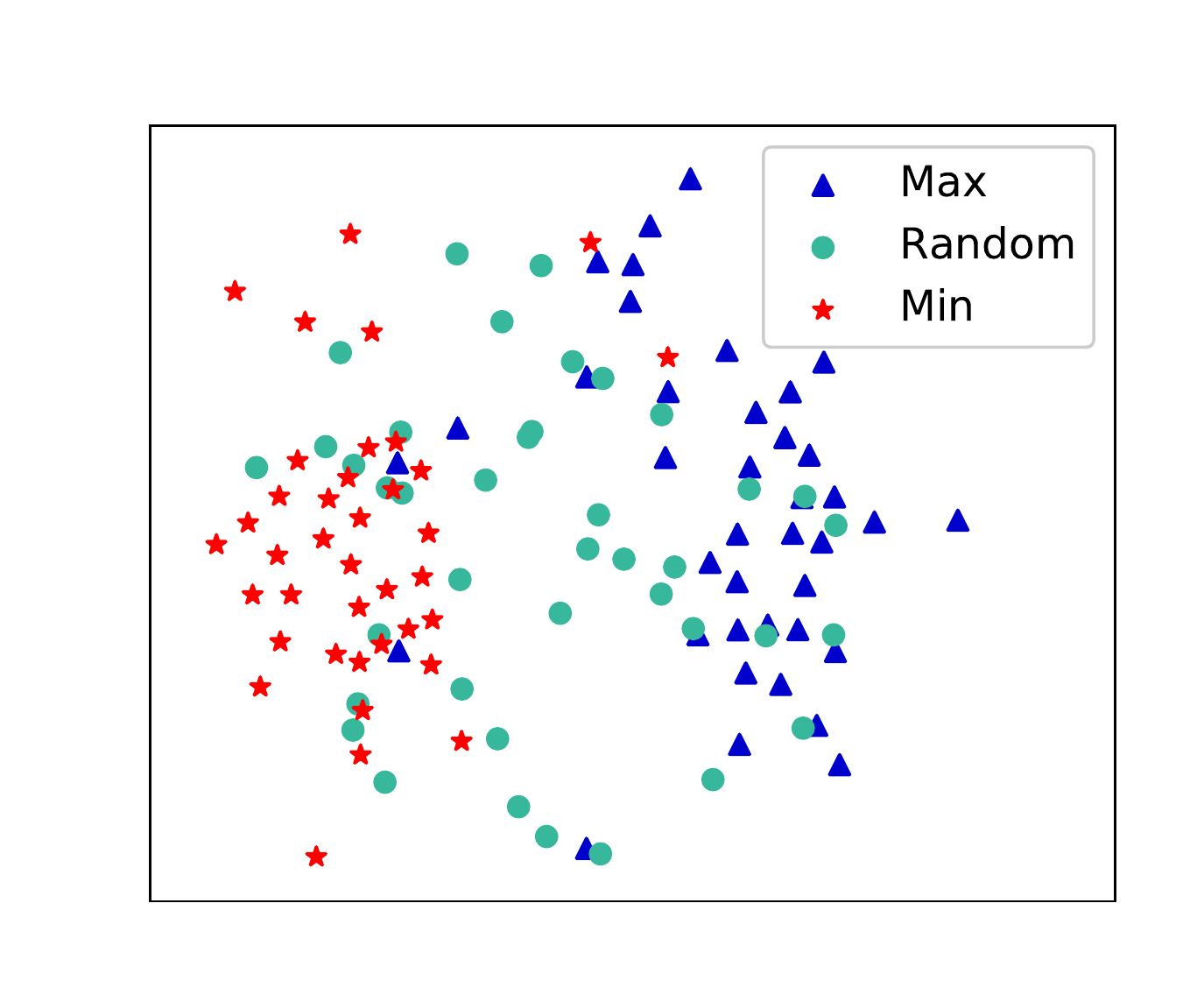}
		\caption{User Embedding}
		\label{fig:ex_user_emb}
	\end{subfigure}
	% \vspace{-0.15cm}
	\caption{Recommendation results and user embeddings for different user groups.}
	% \vspace{-0.3cm}
	\label{fig:explain_experiment}
\end{figure*}

We firstly validate whether the attention mechanism actually play a role for identifying users' preferences. Specifically, according to the learned parameters previously, we select 40 users with the largest actor-attention score, 40 users with the smallest actor-attention and another 40 random users as three separate test groups, and we denote them as \emph{Max, Min}, and \emph{Random} respectively. We then remove the actor network to train a new model, and other settings remain the same. We can then test the recommendation performances for the three test user groups, with the results shown in Fig.~\ref{fig:ex_precision} and Fig.~\ref{fig:ex_ndcg}. 
% We can simply server user embeddings as favorable features for clustering to get a set of users with similar preferences. 
We can see the lack of actor network affects differently on the three test groups. The overall performance order is as follows: {\it Min} > {\it Random} > {\it Max}, showing that \emph{Max} group is severely influenced, which confirms our analysis that these users are more concerned about actors.
Also, the derived user embeddings projected by t-SNE ~\cite{tsne} are also illustrated in Fig.~\ref{fig:ex_user_emb}. It's not hard to find out that {\it Max} and {\it Min} are clearly separated apart from each other, which demonstrates the effectiveness of the learned user embeddings in distinguishing users with different preferences.

In addition, we pick two cases to explain the attention scores output by eRAN in Table~\ref{tab:case_study}. We can see user 255 and user 639 both watch the movie \emph{Fear and Loathing in Las Vegas}. However, the model can distinguish that user 255 is driven by the same actor, while user 639 is driven by the same director according to attention scores. Meanwhile we can compute the most similar movies to them and find that user 225 is keen on the actor \emph{Johnny Depp} while user 639 like the director \emph{Terry Gilliam}.
Therefore, we can easily use eRAN to provide the explanations like {\it "A is similar to B and C, especially with the same {\bf actors}"}.

\begin{figure*}[t!]
	\centering
	\begin{subfigure}[b]{0.48\textwidth}
		\includegraphics[width=\textwidth]{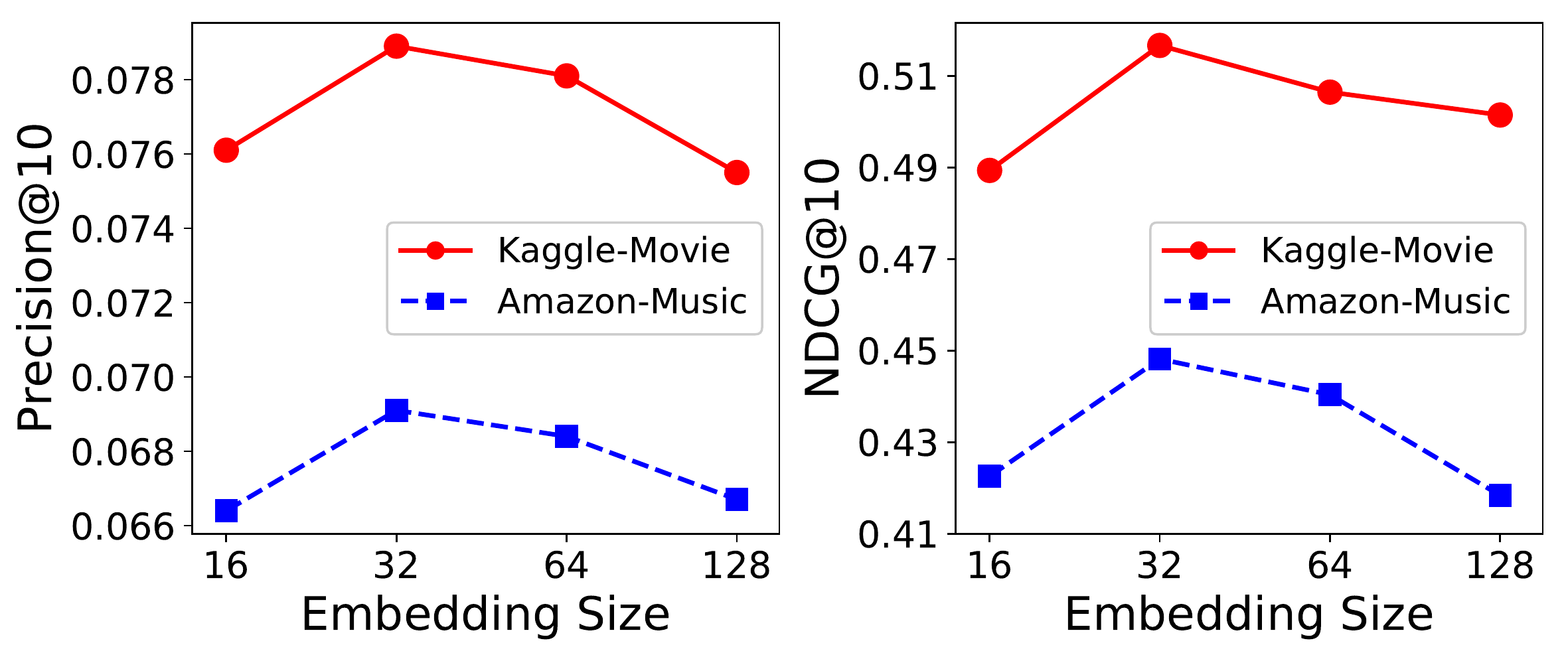}
		\caption{Embedding Size}
		\label{fig:se_embed}
	\end{subfigure}
	%	\hspace{0.005\textwidth}
	\begin{subfigure}[b]{0.48\textwidth}
		\includegraphics[width=\textwidth]{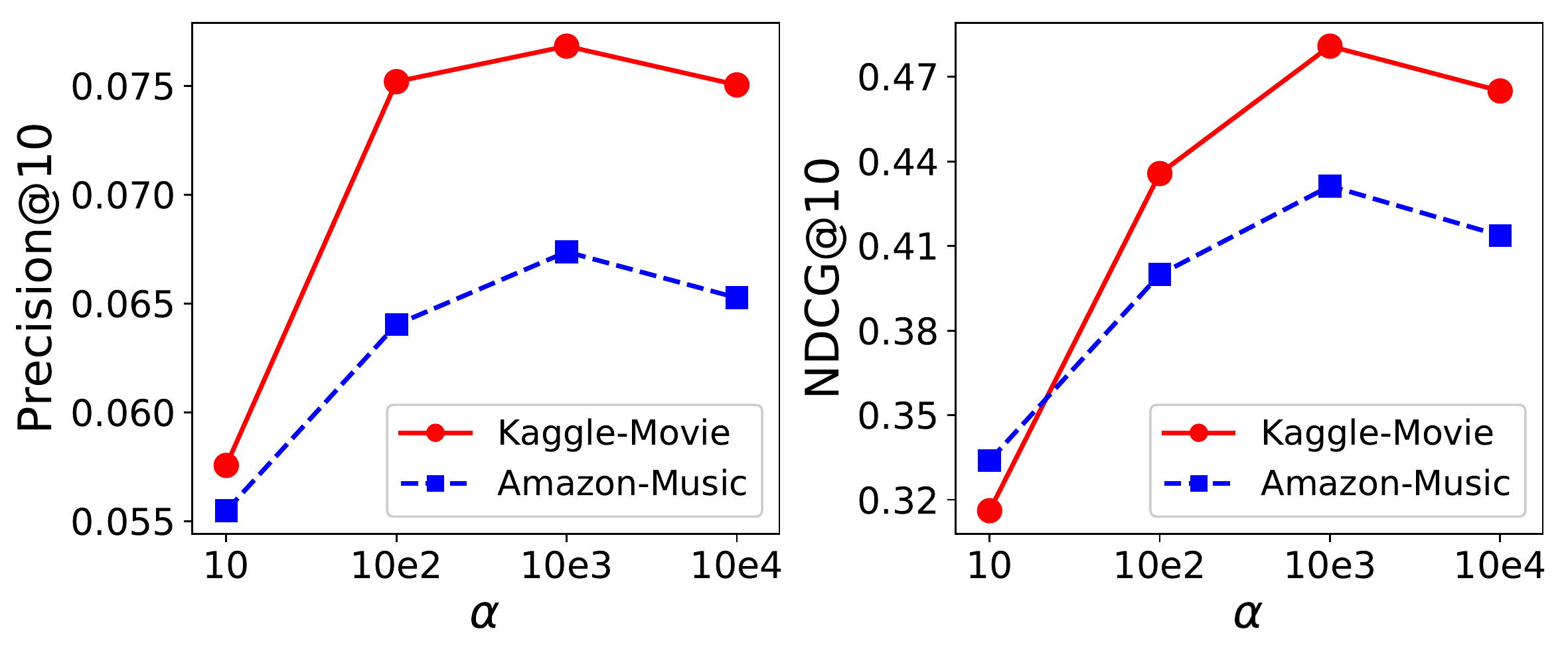}
		\caption{$\alpha$}
		\label{fig:se_reg}
	\end{subfigure}
	% \vspace{-0.15cm}
	\caption{Impact of hyper-parameters on ranking performance.}
	% \vspace{-0.3cm}
	\label{fig:sense_all}
\end{figure*}

\subsection{Parameter Sensitivity}
In this subsection, we examine the sensitivity of two parameters, \emph{i.e.}, the embedding size and the weighting parameter $\alpha$ in the loss function. 
%For the embedding size, we vary it in the set of \{16, 32, 64, 128\}. And for the hyper-parameter $\alpha$, we vary it in the set of \{10, 100, 1000, 10000\}. 

\textbf{Embedding size.} Figure~\ref{fig:se_embed} demonstrates the impact of embedding sizes on the results. It's easy to find that 32 is the best embedding size for both {\it Kaggle-Movie} and {\it Amazon-Music} as measured by Precision and nDCG. Moreover, the performances remain stable in all the settings, which shows the robustness of our method.

\textbf{The weighting parameter $\alpha$.} From the results shown in Figure~\ref{fig:se_reg}, we can see the performance first increases along with $\alpha$, and then begins to drop when $\alpha$ > 1000. It is worth mentioning that our model would reduce to eRAN-L2 when $\alpha$ is infinitesimal, and to eRAN-L1 when $\alpha$ is infinity. Their performances are consistent with the trend indicated by the sensitivity analysis.

\section{Related Work}
Our work is related to the following streams of recommendation methods including item-based, attribute based, as well as explainable recommendation.

%\subsection{Item based recommendation}
The idea of item-based CF methods is that the prediction of a user on a target item depends on the similarity of this item to all items the user has interacted with in the past. Traditional item-based CF methods often predefine some similarity measures such as cosine similarity and Pearson coefficient~\cite{sarwar2001item}. Another common approach is to employ random walks on the user-item bipartite graph~\cite{liu2017related}. However, such heuristic similarity measurement lacks optimization tailored for different datasets, and thus may yield suboptimal results. Recently, Ning et al. has proposed a method SLIM  which learns item similarity directly from data~\cite{ning2011slim}. The idea is to reconstruct the original user-item interaction matrix by the item-based CF model. Afterwards, Kabbur et al. further proposes FISM to explore the low-rank property of the learned similarity matrix to handle data sparsity problem~\cite{kabbur2013fism}. While FISM is shown to outperform recommendation approaches, it has the limitation in estimating only a single global metric for all users. To that end, GLSLIM clusters the users and estimates an independent SLIM model for every user subset~\cite{christakopoulou2016local}, whereas the number of clusters is difficult to determine, and thus the modeling of personalized preferences is coarse-grained.

%As demonstrated in ~\cite{davidson2010youtube,bayer2017generic}, due to the fact that the item neighborhood is sparse, which makes item-based CF methods very suitable for real-time recommendation,  And the other characteristic is that these methods depend heavily on the design of similarity metric. 

%\subsection{Feature based recommendation}
In addition to user-item interactions, many researchers attempt to leverage additional information for recommendation, such as user-item attributes and context information~\cite{zhao2016connecting,baltrunas2011matrix}. FM is an earlier general feature-based framework for recommendation, which is suitable for sparse structured data~\cite{rendle2010factorization}, and it is recognized as the most effective linear embedding method. Due to the recent huge success of deep learning in many fields, some deep variants of FM have been proposed to enhance the model's representation capacity, including AMF ~\cite{xiao2017attentional}, DeepFM ~\cite{guo2017deepfm} and PNN ~\cite{qu2016product}. In these methods, the feature weights are the same for all the users and they cannot capture the fine-grained user preferences, which are not explainable.

%On the other hand, thanks to the powerful feature extraction ability of deep learning, some researchers have explored the influence of some unstructured data on recommendations. For example, DeepCoNN employs CNN to learn low rank vectors from text data ~\cite{zheng2017joint}. VBPR uses CNN to extract image features ~\cite{he2016vbpr} and HLDBN applies DBN to process audio data ~\cite{wang2014improving}. Network is a special kind of unstructured data and there are all kinds of methods to extract features from networks, such as Deepwalk ~\cite{perozzi2014deepwalk} and SDNE ~\cite{wang2016structural}. However, this type of data especially item networks has rarely been used in recommendation field. Therefore, in this paper we construct item networks to fill in this gap.

%\subsection{Explainable Recommendation}
Recently employing auxiliary information to help understand user behaviors and provide explainable recommendations have become prevailing in the research field. Zhang et al. propose EFM ~\cite{zhang2014explicit}, where the basic idea is to align each latent dimension in matrix factorization with a particular explicit feature, and recommend items that performs well on the features that users care about. Chen et al. further extended the EFM model to tensor factorization afterwards ~\cite{chen2016learning}. On the other hand, McAuley and Leskovec propose HFT to understand the hidden factors in latent factor models based on the hidden topics extracted from textual reviews ~\cite{mcauley2013hidden}. After that, many probabilistic graphic model based methods have been proposed for explainable recommendation ~\cite{wu2015flame,ren2017social}. Recently, deep learning and attention mechanism have attracted much attention in the recommendation field, and they have also been wildly applied for explainable recommendations. For example, Seo et al. leverage attention mechanisms upon the user/item reviews to explore the usefulness of reviews and with the learned attention weights, the model can indicate which part is more important ~\cite{seo2017interpretable}. Chen et al. propose VER which can highlight the image regions that a user may be interested in as explanations ~\cite{chen2018visually}. Our work follows this thread but mainly focuses on learning explanations through user behavior data rather than text data.

\section{Conclusions}
\label{sec:conc}
In this paper, we propose a personalized item to item recommendation method eRAN. By formulating the co-purchased relationships and item attributes as multiple \emph{attribute networks}, eRAN combines both views of recommendations. By plugging an attention mechanism in obtaining personalized item representation, eRAN gains ability to derive the attractions of attributes to users and personalized item similarity simultaneously. Experiments on real-world datasets demonstrate the superiority of our methods for recommendation tasks and cold-start items. Moreover, the learned user embeddings and attention weights capture the fine-grained user preferences on attribute level and guide the explanations for recommendations. Future work includes integrating multi-item relationships such as complementation and substitution into our model, and seeking the influence of other different attention mechanisms.

%\section{Acknowledgments}
%Dr. Junjie Wu's work was partially supported by the National Natural Science Foundation of China (NSFC) (71531001, U1636210, 71725002). Dr. Guannan Liu's work was supported in part by NSFC (71701007).

\bibliographystyle{ACM-Reference-Format}
\bibliography{related_works}

\end{document}